\documentclass[%
aip,
amsmath,amssymb,
reprint,%
]{revtex4-1}

\usepackage{graphicx}
\usepackage{dcolumn}
\usepackage{bm}
\usepackage{xcolor}
\usepackage[utf8]{inputenc}
\usepackage[T1]{fontenc}
\usepackage{mathptmx}
\usepackage{hyperref}
\usepackage[all]{hypcap}
\hypersetup{
	colorlinks=true,
	linkcolor=blue,
	filecolor=magenta,      
	urlcolor=cyan,
	pdftitle={Sharelatex Example},
	bookmarks=true,
	pdfpagemode=FullScreen,
}
\begin{document}
	
	\title[]{Effect of Ion Motion on Breaking of Longitudinal Relativistically Strong
		Plasma Waves: Khachatryan mode revisited}
	
	\author{Ratan Kumar Bera}
	\altaffiliation[Present address: The Centre for Space Plasma and Aeronomic Research (CSPAR), University of Alabama Huntsville (UAH),
	Huntsville, AL 35805, USA]{}
	\email{rkb0019@uah.edu}
	
	\author{Arghya Mukherjee}
		\altaffiliation[Present address: Center of Excellence in Space Sciences India, Indian Institute of Science Education and Research Kolkata, Mohanpur - 741246, West Bengal, India]{}
	\affiliation{ 
		Institute for Plasma Research, Bhat, Gandhinagar-382428, India%
	}%

	\author{Sudip Sengupta}
	
	\affiliation{ 
		Institute for Plasma Research, Bhat, Gandhinagar-382428, India%
	}%
	\affiliation{Homi Bhabha National Institute, Training School Complex, Mumbai, 400094, India}
	
	\author{Amita Das}
	\affiliation{%
	    Department of Physics,
		Indian Institute of Technology Delhi, Hauz Khas, New Delhi-110016, India 
	}%
	
	\date{\today}
\begin{abstract}  
Effect of ion motion on the spatio-temporal evolution of a relativistically strong space charge wave, is studied using a 1-D fluid simulation code. In our simulation, these waves are excited in the  wake of a rigid electron beam propagating through a cold homogeneous plasma with a speed close to the speed of light. It is observed that the excited wave is a mode as described by Khachatryan [{\it Phys. Rev. E {\bf 58}, 7799 (1998)}] whose profile gradually sharpens and the wave eventually breaks after several plasma periods exhibiting explosive behaviour. It is found that breaking occurs at amplitudes, which is far below the breaking limit analytically derived by Khachatryan [{\it Phys. Rev. E {\bf 58}, 7799 (1998)}]. This phenomenon of wave breaking, at amplitudes well below the breaking limit, is understood in terms of phase mixing of the excited wave. It is further found that the phase mixing time ( wave breaking time ) scales inversely with the energy density of the wave. 
\end{abstract}

\maketitle 
\section{Introduction}
Over decades, research on relativistically strong plasma waves has attracted
significant amount of attention because of their paramount importance in the progress of 
plasma physics as well as astrophysics. For instance, the excitation and breaking of such
waves serves as a useful paradigm to illustrate the physics of, plasma based 
acceleration schemes \cite{tajima,chen,esarey,uhm,ratan}, 
fast ignition concept in inertial confinement fusion systems \cite{tabak,Pukhov,Wills}, as well as
various solar and astrophysical 
processes \cite{Ferrari,ebisuzaki,shukla,Tsiklauri}. 
Relativistically strong plasma waves (RSWs) were first studied by Akhiezer and Polovin\cite{akhiezer} in their seminal work, where they obtained stationary wave frame solution ( now known as Akhiezer-Polovin mode ) of the relativistic fluid-Maxwell set of equations in the limit of immobile ions. It was shown that the amplitude of a longitudinal RSW is limited by the breaking limit $eE_{WB}/m\omega_{pe} c  = \sqrt{2}(\gamma_{ph} - 1)^{1/2}$, where $\omega_{pe}$ is the non-relativistic electron plasma frequency and $\gamma_{ph} = 1/\sqrt{1 - (v_{ph}/c)^{2}}$ is the Lorentz factor associated with the phase velocity $v_{ph}$ of the wave. With immobile ions, longitudinal Akhiezer-Polovin mode is found to be excited in the wake of rigid electron beam propgating through a cold ummagnetized homogeneous plasma\cite{ratan2}. Effect of ion motion on such a wake wave was investigated by Rosenzweig\cite{rosenion}. A more detailed study of the effect of inclusion of ion dynamics on the longitudinal Akhiezer-Polovin mode was done by Khachatryan\cite{khacha}, and it led to a modified staionary wave frame solution, hereinafter called the ``Khachatryan mode''. Khachatryan\cite{khacha} further showed that ion motion modifies the breaking limit of a longitudinal RSW  as $eE_{WB}/m\omega_{pe} c  = \sqrt{2}\gamma_{ph}[1 + (1 - \xi_{1}^\frac{1}{2}\xi_{2}^\frac{1}{2})/\mu]$, where $\xi_{1} = 1 + \mu$, $\xi_{2} = 1 + [\mu(\gamma_{ph} - 1)/(\gamma_{ph} + 1)]$ and $\mu = m_{e}/m_{i}$ is the electron to ion mass ratio.\\

Effect of ion motion on RSWs can be crucial for understanding some phenomena in laboratory and astrophysical plasmas. For 
example, relativistic electron-positron plasmas and pair ion plasmas
(plasmas consisting of two classes of particles with opposite sign of the charges, but equal masses),
are believed to be present in many extreme astrophysical environments, 
such as Active Galactic Nuclei (AGN), Pulsar Wind  Nebulae (PWN), Gamma  Ray Bursts  (GRBs),  and  Black  Holes  (BHs)
\cite{Hui, Piran, Arons, Hirotani, Waxman}. 
In such scenarios, the excitation of RSWs and their breaking 
are critical for understanding of many astrophysical events like jet formation,
ultra-high-energy-cosmic-rays (UHECRs) generation, shock acceleration process, and many energetic phenomena associated with quasars, GRBs, and BHs
\cite{Ferrari,Hirotani,ebisuzaki, Wardle, Waxman}.
In laboratory experiments, recent reports on plasma based particle acceleration process where RSWs are 
used to accelerate charge particles to high energies, indicate that  the motion of ions can give rise 
to transverse fields that can in turn disrupt the motion of the driver beam and can also effect the energy transfer 
ratio from the driver beam
to the accelerated particles \cite{rosen2,rosenion,viera}.
Therefore, studying the effect of ion motion on the excitation and breaking of RSWs is important for both laboratory and astrophysical plasmas.
Excitation and breaking of longitudinal Akhiezer-Polovin mode
has been recently examined by several authors \cite{prabal,arghya,ratan2}. It is now well established 
that longitudinal Akhiezer-Polovin mode can break much below its wave breaking limit via the process of phase mixing, if
it is subjected to an arbitrarily small amplitude longitudinal perturbation.
The analytical scaling of phase mixing time ( wave breaking time ) with the parameters of the longitudinal Akhiezer-Polovin mode ({\it viz.} phase velocity $\beta$ and fluid velocity amplitude $u_m$) has been derived \cite{arghya} and has also been verified with numerical simulations\cite{arghya,ratan2}. 
To the best of our knowledge, such a study for the  excitation and breaking of ``Khachatryan'' mode has never been conducted, which is the purpose of the present paper.\\

In this paper, using two fluid description we have thoroughly investigated  the spatio-temporal evolution of ``Khachatryan mode'' 
in a cold homogeneous plasma
using a 1-D fluid simulation code. Following the method used in refs.\cite{ratan, ratan2} to excite Akhiezer-Polovin mode, to excite the ``Khachatryan'' mode, here
we have used a rigid homogeneous
pulsed electron beam which propagates
inside a cold unmagnetized homogeneous plasma ( having finite ion mass ) with a speed
close to the speed of light.
As mentioned above, dynamics of both plasma electrons and ions contribute to the excitation of the ``Khachatryan'' mode.
When an electron beam propels inside the plasma it expels the nearby plasma electrons and 
attracts the ions 
due to the space-charge force. The electron beam displaces the plasma electrons and ions in opposite direction.
As the beam propagates inside the plasma, the displaced plasma electrons as well as
ions try to come back to their 
original position to nullify the charge separation. But due to their inertia, they overshoot their original 
position.
As a result an oscillation or a wave is excited at the wake of the beam
having phase velocity equal to the velocity of the beam. \\

In next section (Section II), we present the basic equations governing the 
excitation and spatio-temporal evolution of relativistically strong longitudinal space charge wave driven by a rigid relativistic electron beam propagating through a cold homogeneous plasma with a speed close to the speed of light.
In section III, we briefly discuss our numerical techniques and present our numerical observations along with a detailed discussion of the results. 
Finally a brief summary of our work is presented in section IV.
\section{Governing Equations}
%
The basic equations governing the excitation of longitudinal space charge waves driven by a rigid electron beam propagating with relativistic speeds
through a cold plasma are the relativistic fluid-Maxwell equations. These are the continuity and momentum equations for plasma electrons and ions, and the Poisson's equation. Taking $z$ as the direction of propagation of the beam, the basic equations in normalized form are given by
\begin{equation}
  \frac{\partial n_e}{\partial t}+\frac{\partial (n_e v_e) }{\partial z}=0	\label{pl_cont}
 \end{equation}
 \begin{equation}
  \frac{\partial p_e }{\partial t}+v_e\frac{\partial p_e}{\partial z}=-E      \label{pl_mom}
 \end{equation}
 \begin{equation}
  \frac{\partial n_i}{\partial t}+\frac{\partial (n_i v_i) }{\partial z}=0	\label{ion_cont}
 \end{equation}
 \begin{equation}
  \frac{\partial p_i }{\partial t}+v_i \frac{\partial p_i}{\partial z}=\mu  E     \label{ion_mom}
 \end{equation}
 \begin{equation}
\frac{\partial E }{\partial z}=(n_i-n_e-n_b) \label{pois}
 \end{equation}
where $p_e=\gamma_e v_e$, $p_i=\gamma_i v_i$, $\mu=m_e/m_i$ ( electron to ion mass ratio ) and other symbols have their usual meaning. The above equations have been written using the following normalizations,
$t \rightarrow \omega_{pe}t$, $z \rightarrow \frac{\omega_{pe}z}{c}$, $E \rightarrow \frac{eE}{m_e \omega_{pe} c}$,
$v_e\rightarrow \frac{v_e}{c}$, $v_i\rightarrow \frac{v_i}{c}$, 
$p_e \rightarrow \frac{p_e}{m_e c}$, $p_i \rightarrow \frac{p_i}{m_e c}$, 
$n_e\rightarrow \frac{n_e}{n_0}$, $n_i\rightarrow \frac{n_i}{n_0}$ 
and $n_b\rightarrow \frac{n_b}{n_0}$.
Here $\omega_{pe}$, $n_0$, and $c$ are the electron plasma frequency, equilibrium plasma density and speed 
of light respectively. Beam evolution equations ( beam continuity and momentum ) have been omitted, as we have used a rigid beam to excite the wake waves\cite{ratan2,rosenzweig}. An ideal rigid beam can propagate inside the plasma without any deformation.
In reality this is true only for a sufficiently energetic beam.
In our earlier works \cite{ratan,ratan2}, we have shown that a beam can propagate inside the plasma with negligible deformation ( and can be considered to be rigid ) for hundreds of plasma periods provided the velocity of the beam $v_b \geq 0.99$. In this limit, the evolution equations for the beam can be neglected.
\section{Fluid Simulation, Results and discussion}
In this section, we first briefly describe the numerical techniques used to study the excitation and breaking of longitudinal RSW in a cold plasma. We have developed a 1-D fluid code using LCPFCT set of subroutines which are based on flux-corrected transport (FCT) scheme \cite{boris}. The FCT scheme is basically a generalization of the two-step Lax-Wendroff method\cite{numr}. LCPFCT subroutines are used to solve generalized continuity like equations ( {\it i.e.} equations (\ref{pl_cont}) - (\ref{ion_mom}) ). Poisson's equation (\ref{pois}) is solved using successive over-relaxation (SOR) method and is coupled to the LCPFCT subroutines. Coupling these schemes in a time-centred way, a full 1-D fluid code is developed, which is then used to  solve equations (\ref{pl_cont}) - (\ref{pois}). \\

In the simulation,
the driver beam is allowed to propagate from one end to the 
other end of the simulation box along $z$-direction.
For a given beam profile, we have initialized the simulation using the corresponding analytical profiles of 
plasma electron density, ion density, electron velocity, ion velocity, and electric field as
given by Rosenzweig et al. \cite{rosenion} and then numerically followed the space-time evolution of the system using equations (\ref{pl_cont}) - (\ref{pois}). 
The results obtained from simulation are checked by repeating the simulations for different mesh sizes. 
In the following subsections, we discuss the simulation results in detail.
%
%
\subsection{Excitation of longitudinal relativistically strong plasma waves }
\begin{figure}[htb!]
	\includegraphics[width=0.5\textwidth]{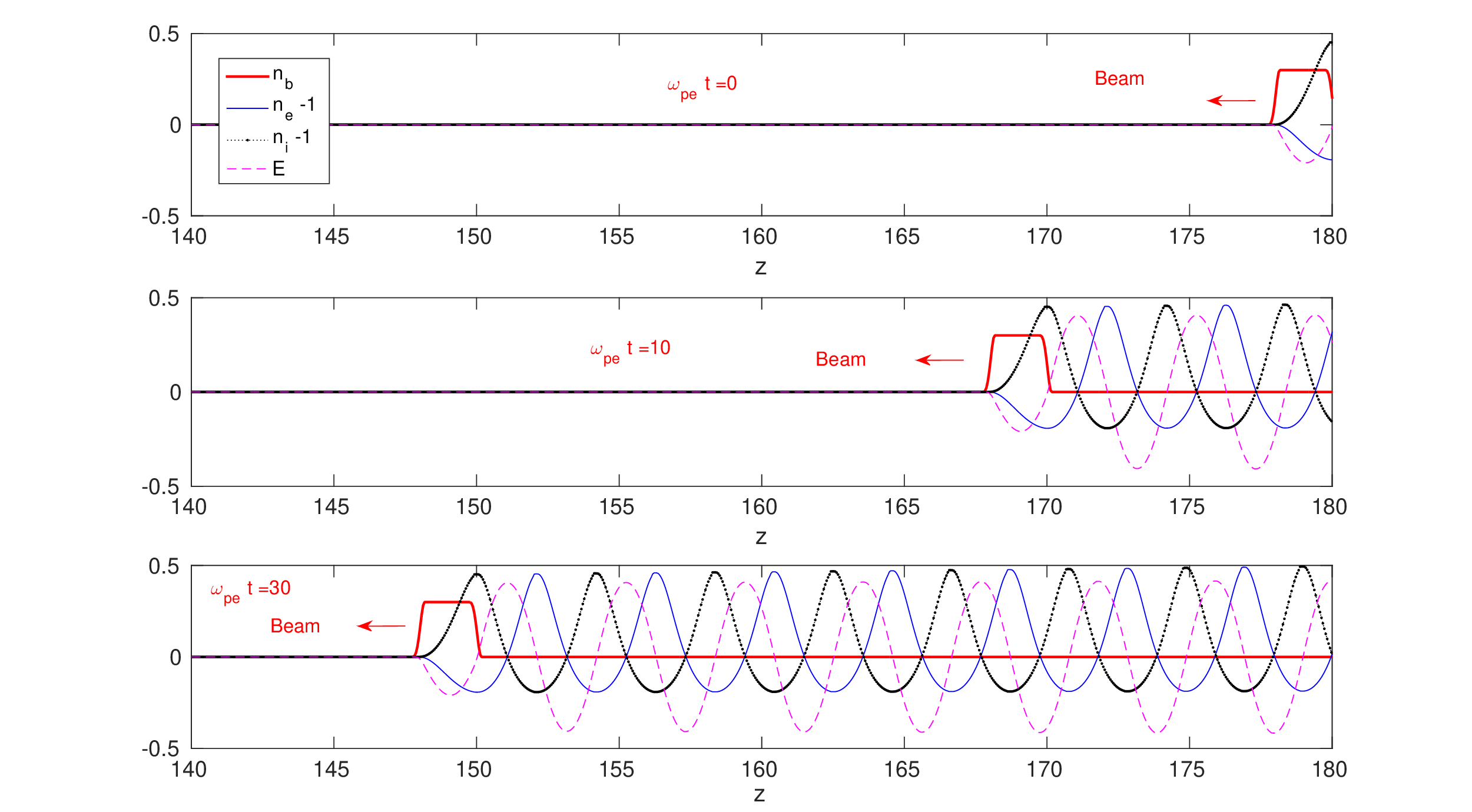}
	\caption{Plot of normalized perturbed electron density ($n_e -1$),  ion density  ($n_i -1$), and electric field ($E$)
		profiles at different times for normalized beam density ($n_b = 0.3$), beam length ($l_b = 4$), beam velocity ($v_b
		= 0.9999$) and mass ratio ($\mu = 1$)}
	\label{fig1}
\end{figure}

\begin{figure}[htb!]
	\includegraphics[width=0.5\textwidth]{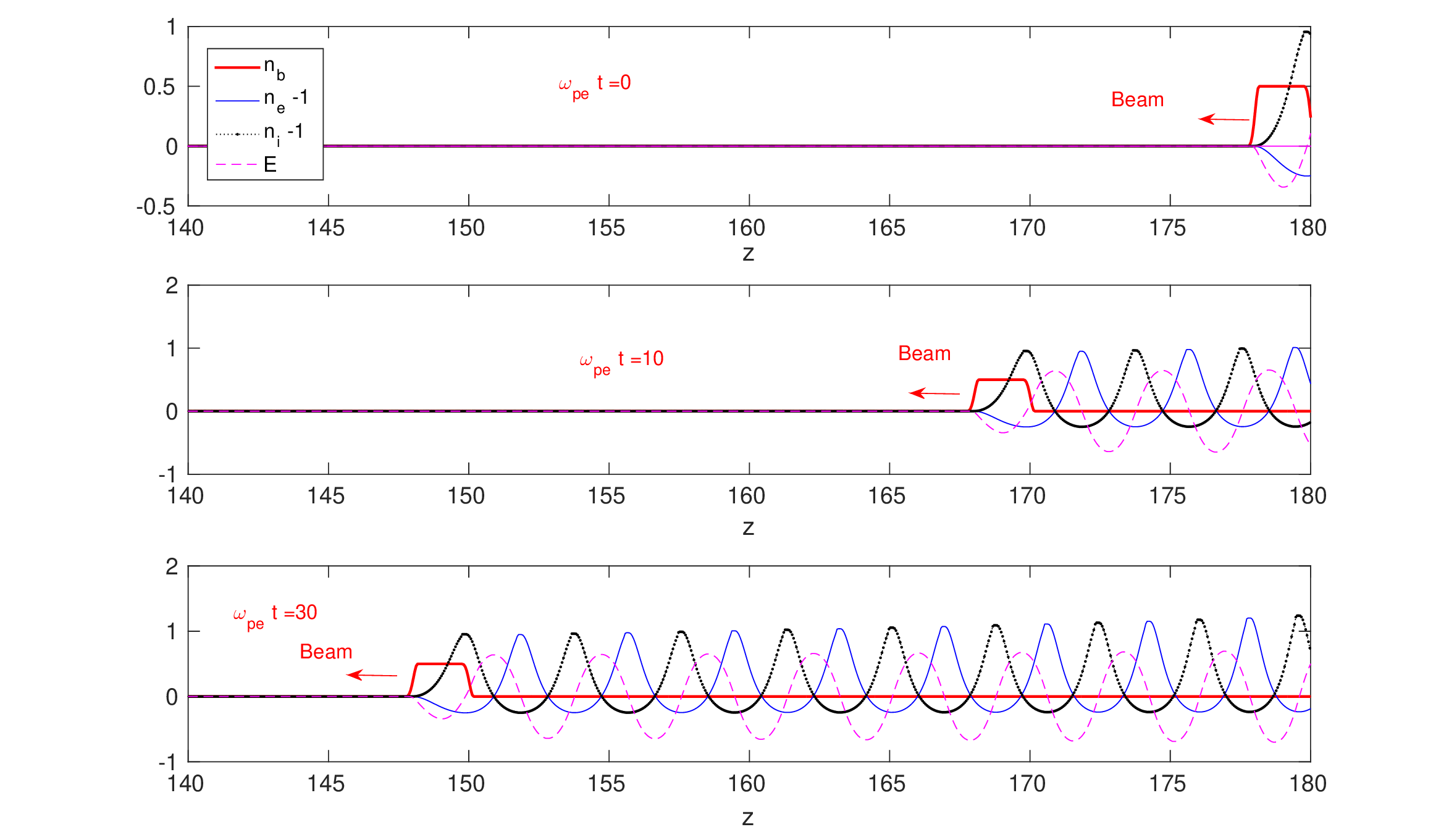}
	\caption{Plot of normalized perturbed electron density ($n_e -1$),  ion density  ($n_i -1$), and electric field ($E$)
		profiles at different times for normalized beam density ($n_b =0.5$), beam length ($l_b = 4 $), beam velocity ($v_b = 0.9999$)
		and mass ratio ($\mu = 1$) }
	\label{fig2}
\end{figure}

\begin{figure} [htb!]
	\includegraphics[width=0.5\textwidth, height=0.3\textwidth]{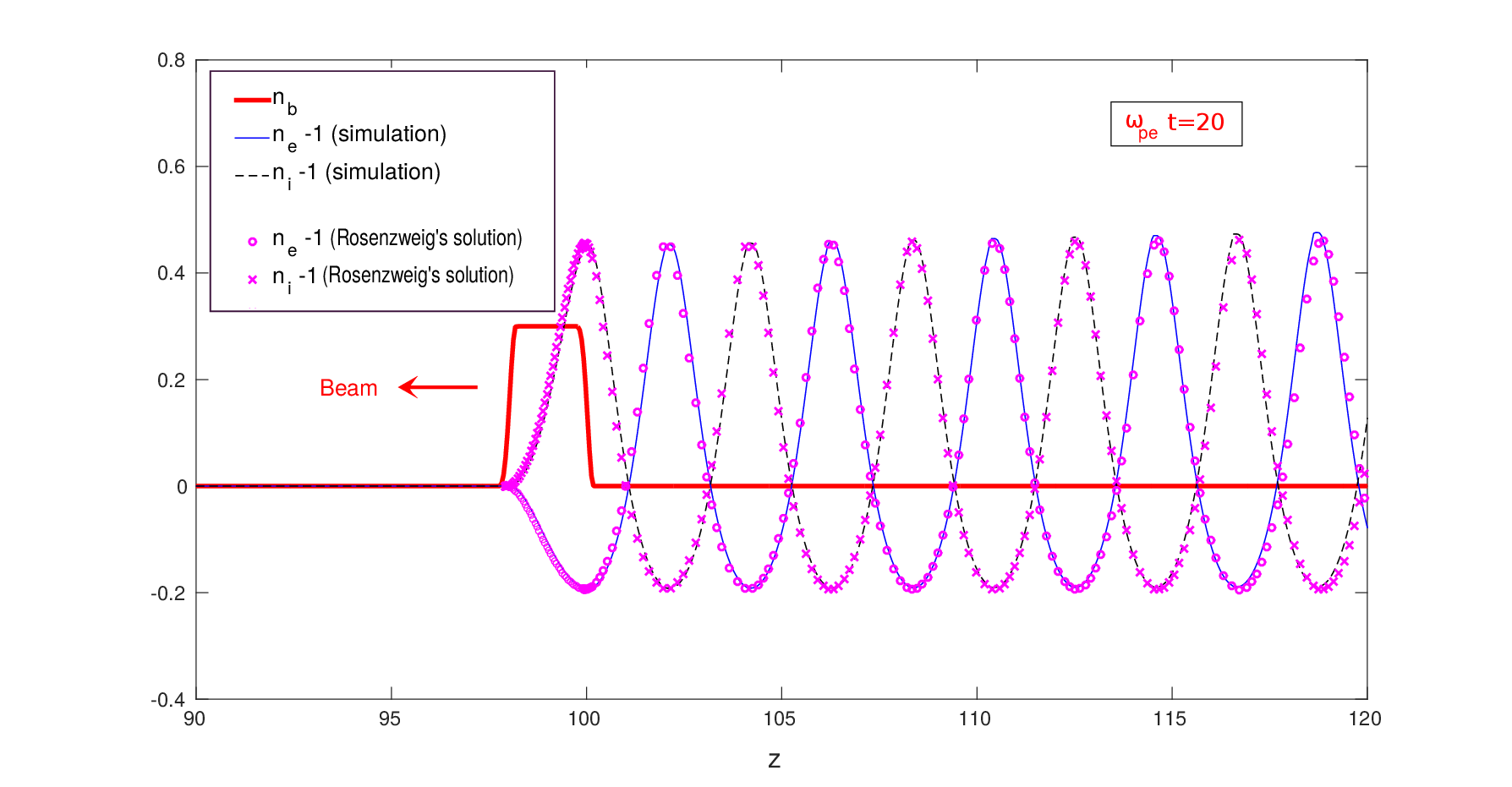}
	\caption{Plot of numerical and analytical normalized perturbed electron density ($n_e -1$) and ion density ($n_i -1$)
		profiles at $\omega_{pe} t =20$ for normalized beam density ($n_b = 0.3$), beam length ($l_b = 4$),
		beam velocity ($v_b = 0.9999$) and mass ratio ($\mu = 1$)}
	
	\label{fig3}
\end{figure}

%
We now present simulation results showing excitation of longitudinal  relativistically strong plasma waves in a cold unmagnetized homogeneous plasma using a rigid electron beam. The simulations have been performed for different beam densities ($n_b$) and mass ratios ($\mu$).
In all the simulation runs, the beam velocity is kept fixed at $v_b=0.9999$ and beam length at $l_b=4$. 
Hence the phase velocity ($v_{ph}$) of the excited wake wave is fixed at $0.9999$. For a fixed plasma density $n_0$,
the amplitude and frequency of excited wave is now completely determined by the values of $n_b$ and $\mu$.
By changing the values of $n_b$ and $\mu$ one can excite wake waves of different amplitude and frequency.
In Figs. (\ref{fig1}) and (\ref{fig2}), for $\mu = 1$, we have plotted
the perturbed plasma electron density ($n_e -1$), ion density ($n_i -1$), and electric field ($E$) profiles at different times $\omega_{pe}t = 0, 10$, and $30$ as obtained from simulation, for $n_b = 0.3$ and $n_b = 0.5$ respectively. 
To initiate the simulations at $\omega_{pe}t=0$, in each case, we have used the analytical profiles for $n_e-1$, $n_i-1$, $v_e$, $v_i$, and $E$, which are obtained by solving equations (20-23) of ref. \cite{rosenion} for both inside and outside the beam. 
We see that the beam excites wake wave as it passes through the plasma.
It is also seen that the amplitude of the wake wave increases with increasing $n_b$ for a given value of $\mu$.  Accuracy of our simulation can be easily judged from Fig. (\ref{fig3}) where we have plotted the numerical profiles at $\omega_{pe} t = 20 $ for $n_b= 0.3$ and $\mu = 1$ along with the analytical solutions obtained by solving equations (20 - 23) of  ref. \cite{rosenion} both inside and outside the beam. The numerical solutions show a good match with the analytical solutions.
\begin{figure}[htb!]
	\includegraphics[width=0.5\textwidth, height=0.3\textwidth]{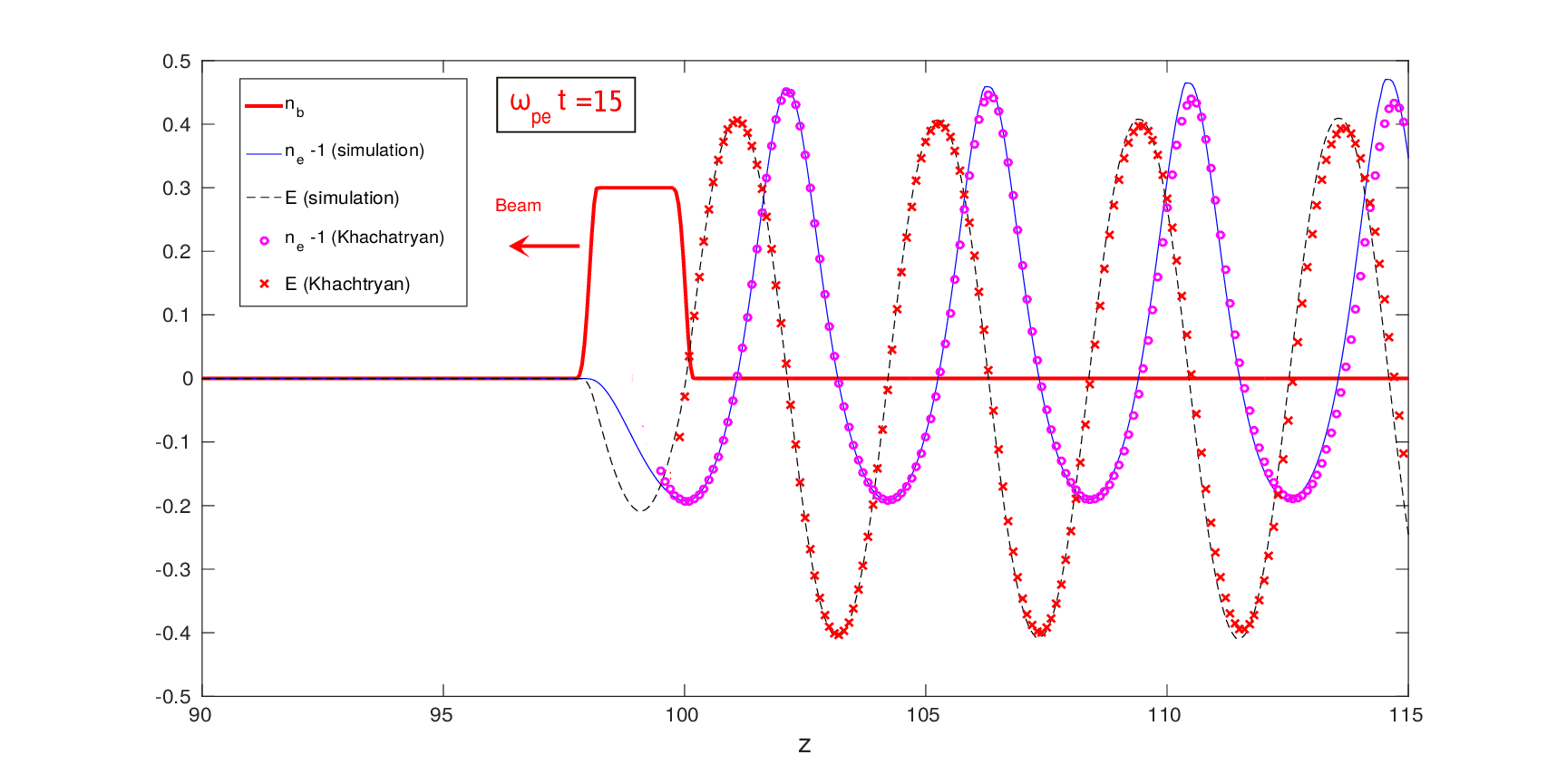}
	\caption{Plot of normalized perturbed electron density ($n_e -1$) and electric field ($E$)
		profiles obtained from simulation and corresponding Khachatrayan mode (analytical) 
		at $\omega_{pe} t =15$ for the normalized beam density ($n_b=0.3$), beam length ($l_b = 4$),
		beam velocity ($v_b =0.9999$) and mass ratio ($\mu =1$).}
	\label{fig4}
\end{figure}
We now note that without the beam (i.e. $n_b = 0$), equations (\ref{pl_cont}) - (\ref{pois}) are the exactly the same equations which were investigated analytically by Khachatryan\cite{khacha} using pseudo-potential method. Stationary wave frame solution of these equations is the ``Khachatryan'' mode, as defined in the introduction. Since the beam density vanishes at the wake of the beam, we expect that 
the wake wave excited by the beam must be a ``Khachatryan'' mode. As discussed in ref.\cite{khacha}, such a mode can be parametrized in terms of $E_{max}$, $v_{ph}$ and $\mu$ where $E_{max}$ represents the electric field amplitude of the excited mode. In Fig. (\ref{fig4}), for $n_b = 0.3$, using the numerical value of $E_{max}$ of the excited wake obtained from simulation, $v_{ph}=v_b$ and $\mu = 1$, we have solved equations (4-9) of ref. \cite{khacha} and plotted the corresponding wave form
on top of the simulated wake wave. We observe a close match between the two. Same exercise has been repeated for $n_b = 0.5$ and $\mu = 1/2000$ (Fig. (\ref{fig5})), again exhibiting a close match between simulation and theory; thus proving that the excited wake wave is a ``Khachatryan'' mode.
\begin{figure}[htb!]
	\includegraphics[width=0.5\textwidth, height=0.3\textwidth]{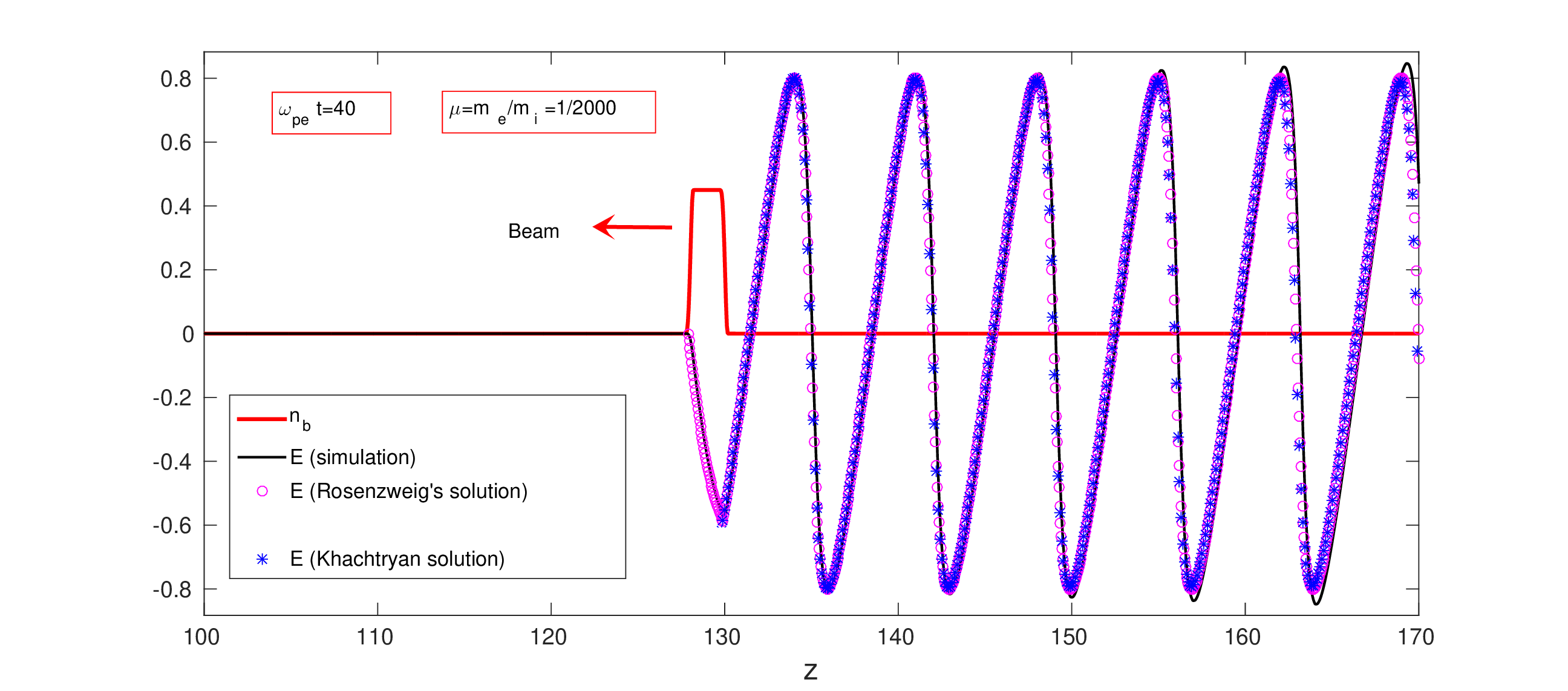}
	\caption{Plot of numerical and analytical profiles of normalized electric field ($E$) 
		for the mass ratio ($\mu = 1/2000$) at $\omega_{pe} t =40$ for the normalized beam density ($n_b=0.5$), beam length ($l_b=4$),
		beam velocity ($v_b =0.9999$).}
	\label{fig5}
\end{figure}

\subsection{Breaking of longitudinal relativistically strong plasma waves}
\begin{figure}[htb!]
	\includegraphics[width=0.5\textwidth, height=0.3\textwidth]{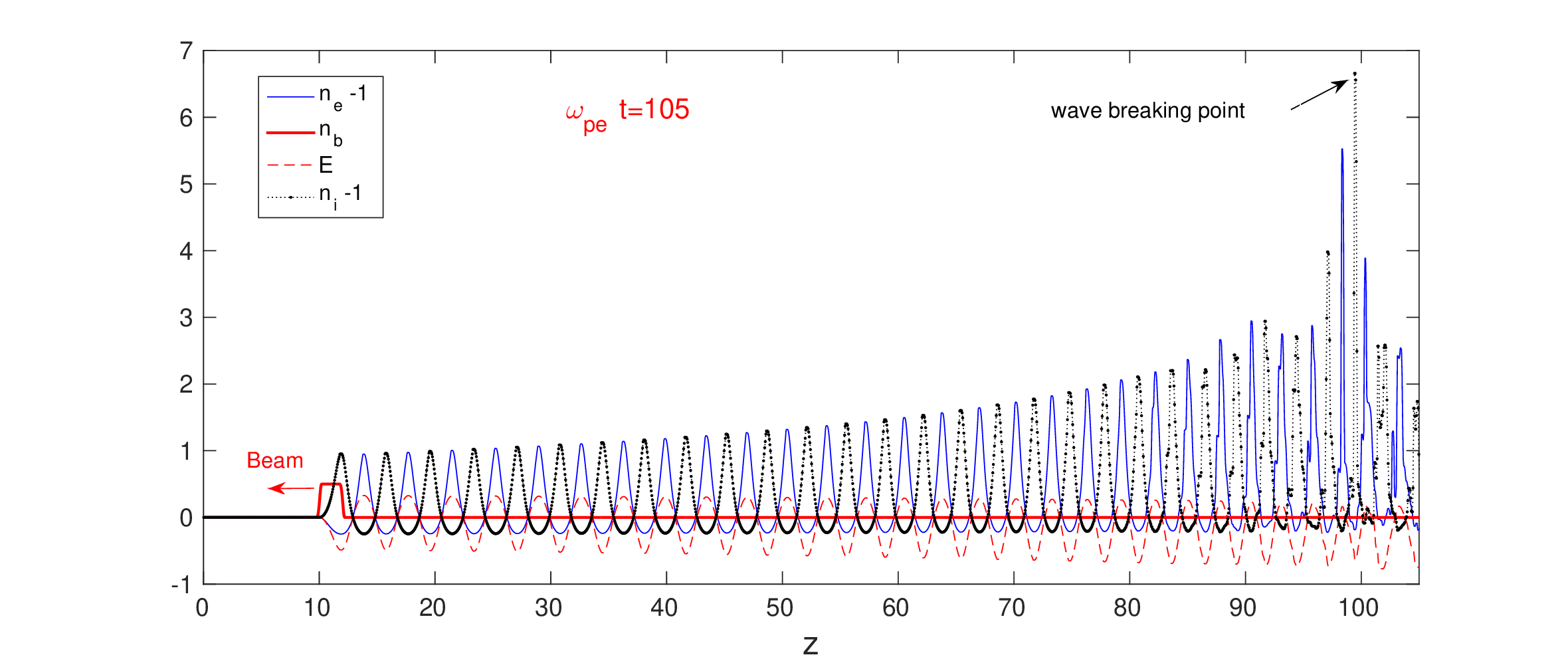}
	\caption{Plot of numerical profile of normalized perturbed electron density ($n_e -1$),  ion density  ($n_e -1$), and electric field ($E$)
		at $\omega_{pe} t =125$ for the normalized beam density ($n_b=0.5$), beam length ($l_b=4$), beam velocity ($v_b
		=0.9999$) and mass ratio ($\mu =1$).}
	\label{fig6}
\end{figure}

\begin{figure}[htb!]
	\includegraphics[width=0.5\textwidth,height=0.3\textwidth]{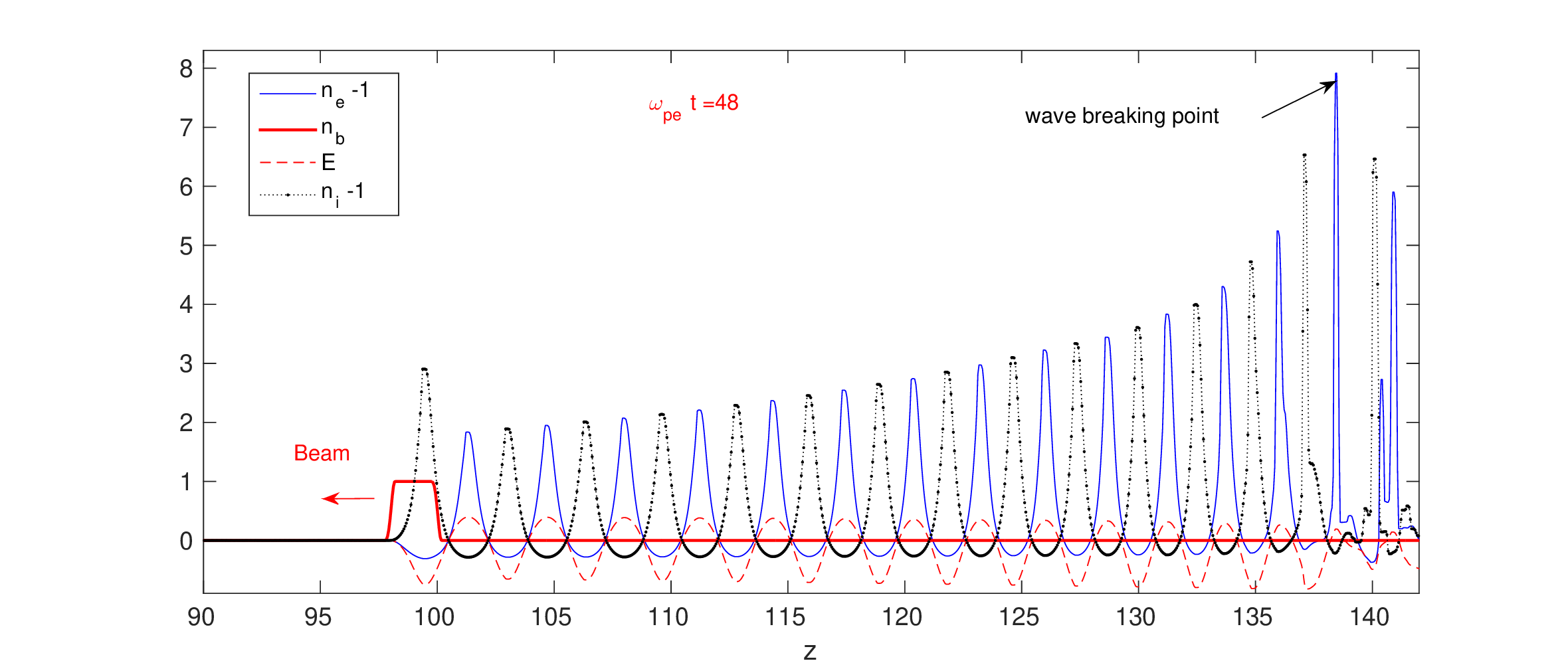}
	\caption{Plot of numerical profile of normalized perturbed electron density ($n_e -1$),  ion density  ($n_e -1$), and electric field ($E$)
		at $\omega_{pe} t =125$ for the normalized beam density ($n_b=1$), beam length ($l_b=4$), beam velocity ($v_b
		=0.9999$) and mass ratio ($\mu =1$).}
	\label{fig7}
\end{figure}

We now numerically follow the spatio-temporal evolution of the wake waves for longer times. It is observed that the profiles of electron density, ion density and electric field gradually changes with time and after several tens of plasma periods, significantly deviates from their analytical profiles. In Figs. (\ref{fig6}) and (\ref{fig7}), we have plotted the profiles of the perturbed electron density ($n_e -1$), ion density ($n_e -1$) and electric field ($E$) for $n_b = 0.5$ at 
$\omega_{pe}t = 105$ and for $n_b = 1$ at $\omega_{pe}t = 48$ respectively.
It is observed that electron density and ion density exhibit ``spiky'' features at later times. The density amplitude ( for electrons and ions ) gradually increases with time and after a certain time
(e.g. $\omega_{pe}t=90$ for $n_b=0.5$ (see Fig. \ref{fig6}) and  
 $\omega_{pe}t=38$ for $n_b=1$ (see Fig. \ref{fig7})), it becomes maximum and then suddenly drops. 
This is a clear signature of wave breaking \cite{sudip_ppcf,sudip_pre,arghya,ratan2,arghya2}. We define the time at which the density ( electron and ion ) amplitude peaks and then suddenly drops, as the wave breaking time. \\

\begin{figure}[htb!]
	\includegraphics[width=0.5\textwidth,height=0.3\textwidth]{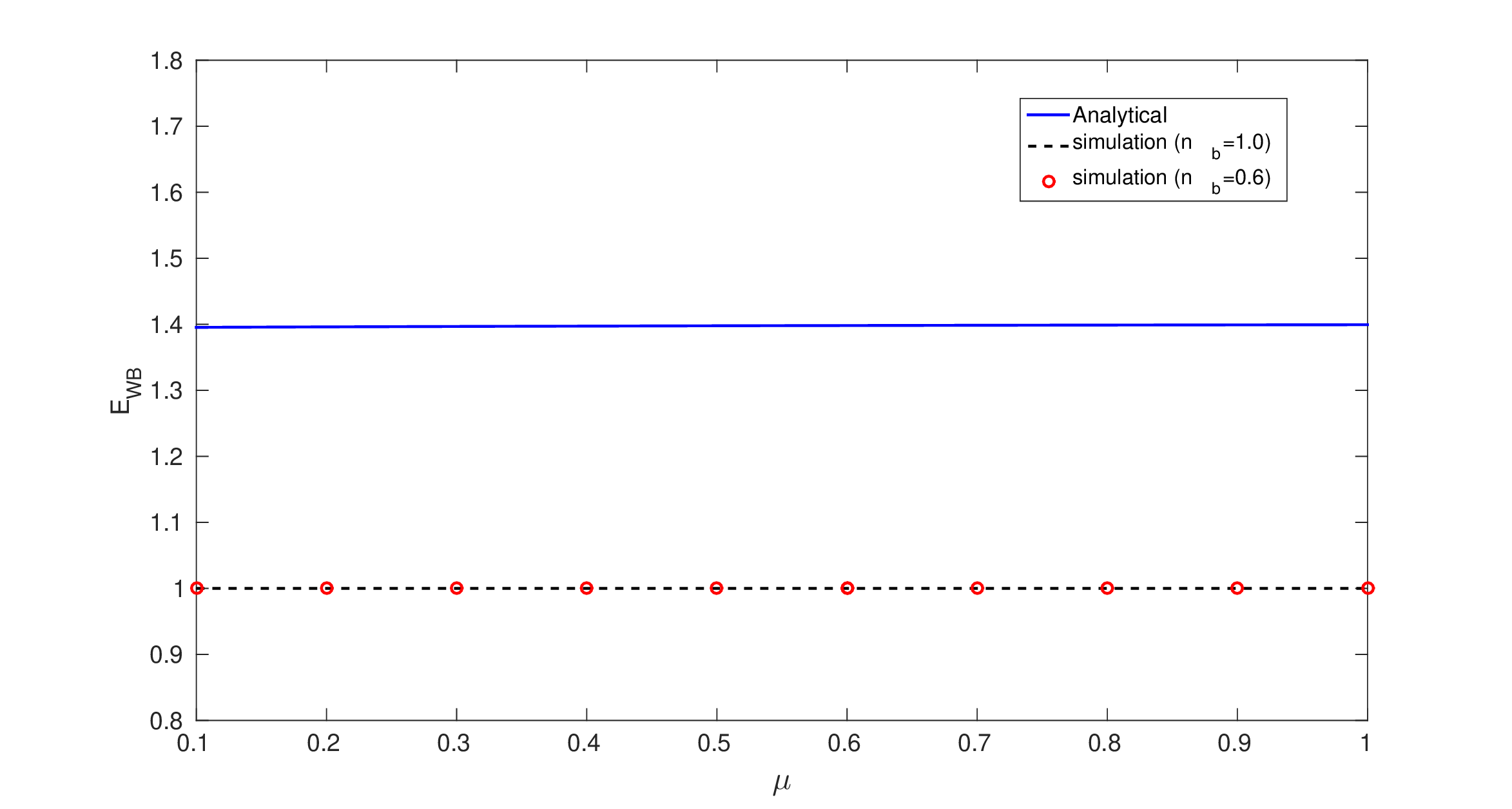}
	\caption{Plot of theoretical and numerical wave breaking electric field ($E_{WB}$)  vs. mass ratios ($\mu$) 
		for normalized beam density $n_b=0.6$ and $n_b=1$ ; where beam length ($l_b$)=4, beam velocity ($v_b$) =0.9999}
	\label{fig8}
\end{figure}

As stated in the introduction, for a ``Khachatryan'' mode,
the wave breaking limit is analytically given as\cite{khacha},
$E_{WB}=\sqrt{2} \gamma_{ph} [1+(1-\xi_1^\frac{1}{2} \xi_2^\frac{1}{2})/\mu]$;
where $\xi_1 =1+\mu$, $\xi_2 =1+[\mu(\gamma_{ph}-1)/(\gamma_{ph}+1)]$ and $\gamma_{ph}=(1-v_{ph}^2)^{-1/2}$. This expression weakly depends on $\mu$ (electron to ion mass ratio); for $\mu = 1$ and $v_{ph} = v_b = 0.9999$, the analytical value of wave breaking limit turns out to be $E_{WB} \sim 1.4$. We have carried out our simulations for two values of driver beam densities ($n_b = 0.6$ and $n_b = 1$) with $v_{b} = 0.9999$ and have recorded the electric field amplitude $E_{WB}$ at the wave breaking time for different values of $\mu$.
Fig. (\ref{fig8}) shows these recorded values as a function of $\mu$ along with their corresponding analytical values. 
We see that the numerical wave breaking limit lies much below the analytical limit, thus indicating that the wave breaks much before it touches the analytical limit.
This difference between the analytical and the numerical breaking limit of relativistically strong plasma waves may be understood in terms of phase mixing process \cite{sudip_ppcf,arghya,ratan2}. 
Phase mixing occurs when the frequency of the wave becomes space dependent\cite{dawson,dawson_kaw}. As a result neighbouring fluid elements or particles sustaining the wave oscillate with different frequencies, which causes the phase difference between them to slowly change with time. Therefore, as time progresses, the profile of the wave gradually gets modified. Eventually a time comes when two neighbouring elements at some spatial location oscillate out of phase, and as 
a consequence they cross each other and the wave breaks exhibiting sharp spikes in the density profile. \\

As stated in the introduction the affect of phase mixing process on the wave-breaking of relativistic electron plasma wave or Akhiezer-Polovin mode has already been  extensively studied by several authors \cite{sudip_pre, sudip_ppcf, prabal, prabal2, arghya}. It is shown that Akhiezer-Polovin mode can break much below its wave breaking limit due to phase mixing process, if it is subjected to a arbitrarily small longitudinal perturbation. As a result of the perturbation, the wave frequency acquires a spatial dependence which leads to phase mixing as described above.
In one of our earlier simulations\cite{ratan2} with immobile ions, 
we have shown that the wake wave excited by a rigid relativistic electron beam propagating through a cold unmagnetized homogeneous plasma, 
is nothing but a Akhiezer-Polovin (AP) mode\cite{akhiezer};
and it breaks much below the analytical wave breaking limit. The longitudinal perturbation in this case arises because of numerical fluctuations which is inherent in any simulation\cite{ratan2}.\\

\begin{figure}[htb!]
\includegraphics[width=0.5\textwidth, height=0.35\textwidth]{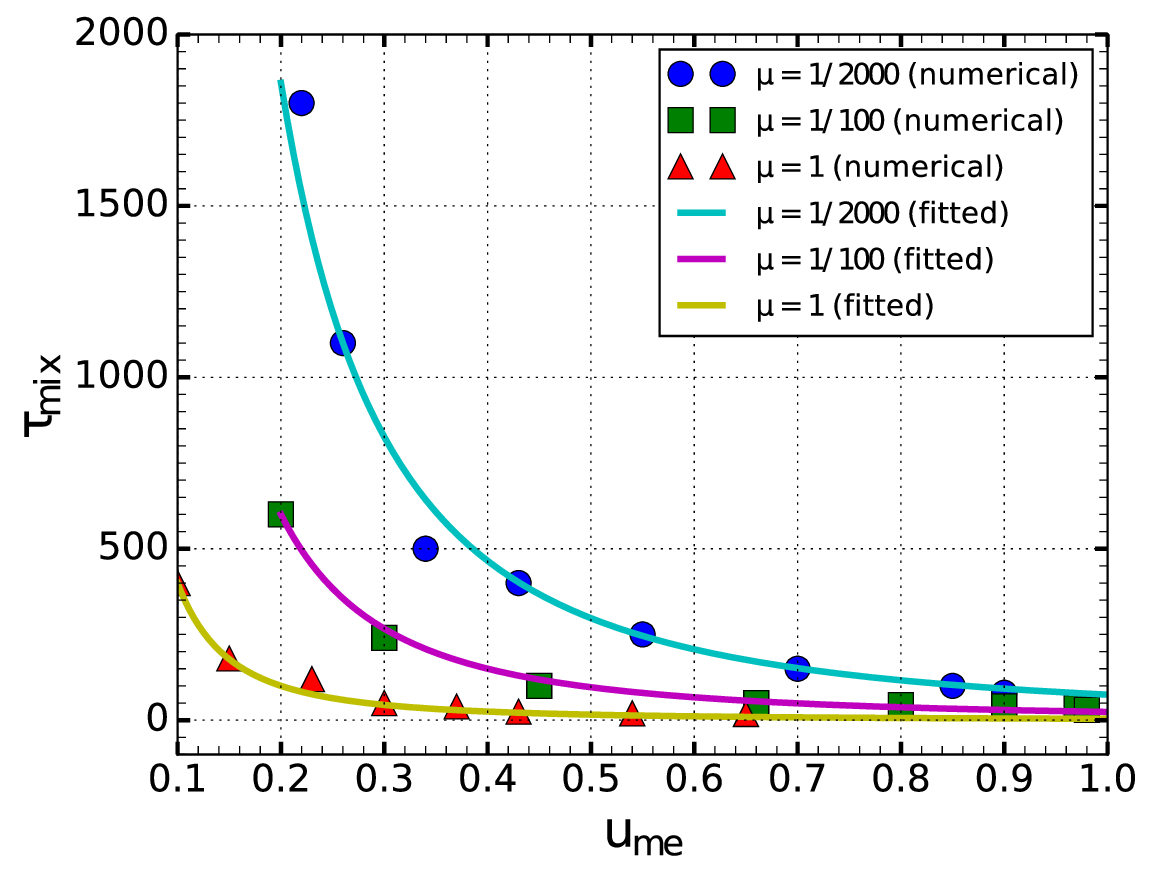}
\caption{Plot of normalized numerical wave breaking time or phase mixing time ($ \tau_{mix}$) obtained from simulation (solid bullet) along with the fitted curve (solid lines) as a function of $u_{me}$; where $u_{me}$ is the maximum electron fluid velocity, for different values of mass ratios ($\mu$).}
\label{fig9}
\end{figure}

Similar observations have been made in the present set of simulations with mobile ions. Due to the presence of mobile ions, though
the basic characteristics of the wave has changed {\it viz.} from Akhiezer-Polovin mode to ``Khachatryan'' mode, but the wave breaking mechanism is found to be the same.
The beam excites a pure Khachatryan mode, as shown in previous subsection in which the frequency is initially independent of space. At later times, the original solution of Khachtryan\cite{khacha} which is  excited by the beam gets perturbed by the inherent numerical fluctuations. As a result, the frequency of the pure ``Khachatryan'' mode becomes
space dependent; this may be seen in Figs. (\ref{fig6}) and (\ref{fig7}), where the distance between the density peaks slowly changes with space. This leads to breaking of the mode via phase mixing process at an amplitude which is much below its wave breaking limit. It is shown in ref.\cite{arghya}, that the  phase mixing time (wave breaking time) for the Akhiezer-Polovin mode, which is supported by electron motion only, scales inversely with the energy density of the electron fluid. Taking a clue from this study, we conjecture that the phase mixing time for the ``Khachatryan'' mode, which is supported by both electron and ion motion, scales inversely with net energy density which includes the energy densities of both the electron and ion fluids {\i.e. $\tau_{mix} \propto 1/[(\gamma_{me} - 1) + ( \gamma_{mi} - 1 )/\mu] \sim 1/[(1 + \mu)u_{me}^2]$}, using $u_{mi} \approx \mu u_{me}$ where $u_{me}$, $u_{mi}$ are respectively the electron and ion velocity amplitude and $\gamma_{me} = 1/\sqrt{1 - u_{me}^2}$, $\gamma_{mi} = 1/\sqrt{1 - u_{mi}^2}$. To verify this conjecture, we have carried out simulations, for fixed $\mu$ and $v_b=0.9999$, with different values of $n_b$. For a given $\mu$ increasing the value of $n_b$ increases the energy density of the electron and ion fluids ( see figures (\ref{fig1})-(\ref{fig2}) ). Fig. (\ref{fig9}) shows the variation of phase mixing time (wave breaking time) with $u_{me}$ for three different values of $\mu = 1, \, 1/100, \, 1/2000$. Here the symbols represent the measured phase mixing time and the continuous line represents our fitting ($\tau_{mix} \sim 1/[(1 + \mu)u_{me}^2]$), which clearly supports our conjecture.
\section{Summary}
In summary, in the present paper, using two fluid description, the excitation and spatio-temporal evolution of relativistic electron beam driven wake wave in a cold homogeneous unmagnetized plasma has been studied using 1-D fluid simulation techniques. The numerical profiles of the wake wave obtained from the fluid simulation
show a good agreement with the corresponding analytical
profiles  obtained from the theory given by Rosenzweig\cite{rosenion} and Khachatryan\cite{khacha}. 
Further space-time evolution of the excited wave shows gradual modification with time and after several plasma periods significantly deviates from the analytical solution of Rosenzweig\cite{rosenion} and Khachatryan\cite{khacha}. 
We observe that the density profile associated with the wake wave becomes spiky at a later time, which is a clear signature of ``wave breaking'' 
\cite{ratan2,arghya,sudip_pre,prabal2,infeld,sudip_ppcf,arghya3}. 
At the point when the wave breaks or density bursts form,
we have recorded the electric field amplitude of the wave.
It is observed that this electric field amplitude ( numerical wave breaking limit )
lies much below than the analytical limit given by Khachatryan \cite{khacha}. 
This difference in the analytical and numerical wave breaking limit has been understood in terms of phase mixing 
process \cite{ratan2,prabal,sudip_pre,arghya}.
It has been shown that the Khachatryan mode breaks much below its analytical wave breaking limit due to the gradual process of phase mixing, which is triggered by numerical fluctuations. Furthermore, it is found that the phase mixing time of the ``Khachatryan'' mode scales inversely with the energy density of the wave. Like previous studies\cite{sudip_pre,arghya4}, study of dependence of phase mixing time of the ``Khachatryan'' mode on amplitude of imposed perturbation and its spectral content is presently under investigation and will be presented elsewhere\cite{arghya3}.
\section{Acknowledgements}
AD would like to acknowledge her J. C. Bose fellowship grant
JCB/2017/000055 and the CRG/2018/000624 grant of
DST for the work.

\section{Data availibility}
The data that support the findings of this study are available
from the corresponding author upon request.

\newpage
\nocite{*}
\bibliography{references}

\end{document}